**Strain-Modulated Interlayer Charge and Energy Transfers in MoS$_2$/WS$_2$ Heterobilayer**

Joon-Seok Kim,[1,2,*] Nikhilesh Maity,[3] Myungsoo Kim,[2] Suyu Fu,[4] Rinkle Juneja,[3] Abhishek K. Singh,[3] Deji Akinwande,[2,*] and Jung-Fu Lin[4,*]

[1] Department of Materials Science and Engineering, Northwestern University, Evanston, IL 60208, USA

[2] Microelectronics Research Center, The University of Texas at Austin, Austin, TX 78758, USA

[3] Materials Research Centre, Indian Institute of Science, Bangalore 560012, India

[4] Department of Geological Sciences, Jackson School of Geosciences, The University of Texas at Austin, Austin TX 78712, USA

**Abstract**

Excitonic properties in 2D heterobilayers are closely governed by charge transfer (CT) and excitonic energy transfer (ET) at van der Waals interfaces. Various means have been employed to modulate the interlayer CT and ET, including electrical gating and modifying interlayer spacing, but with limited extent in their controllability. Here, we report a novel method to modulate these transfers in MoS$_2$/WS$_2$ heterobilayer by applying compressive strain under hydrostatic pressure. Raman and photoluminescence measurements, combined with density functional theory calculations show pressure-enhanced interlayer interaction of the heterobilayer. Photoluminescence enhancement factor η of WS$_2$ in heterobilayer decreases by five times up to ~4 GPa, suggesting a strong ET, whereas it increases by an order of magnitude at higher pressures and reaches almost unity, indicating enhanced CT. Theoretical calculations show that orbital switching in the conduction bands is responsible for the modulation of the transfers. Our findings provide a compelling approach towards effective mechanical control of CT and ET in 2D excitonic devices.

**Introduction**

The ability to couple and control exciton emission is a keystone challenge for realizing excitonic solid state devices, where the computational and transmittance efficiencies are anticipated to be greatly enhanced.[1,2] In the light of the exciton control, two-dimensional (2D) materials are attracting significant interest as an optimal material system, owing to their strong light-matter interaction and large excitonic binding energies.[3,4] Combined with the ability to selectively stack a wide variety of atomic layers, 2D heterostructures possess great potential to be employed in excitonic device applications.[5–8] For example, electrical transport control of the interlayer excitons, where bound electrons and holes reside in the opposite layers, has been achieved in type-II transition metal dichalcogenide (TMDC) heterostructures.[6,7] The successful control of the interlayer excitons further implies the possibility of controlling the excitonic behavior by modulating interlayer charge transfer (CT) and excitonic energy transfer (ET). In this study, we present modulation of the interlayer CT and ET by applying compressive strain to tune the van der Waals (vdW) interaction between $WS_2$ and $MoS_2$ monolayers under hydrostatic pressures. Our combined experimental and theoretical results show that the interlayer interaction is enhanced as the normal compressive strain is applied, leading to an effective modulation of the CT and ET. As a result, the photoluminescence (PL) enhancement factor was tuned by more than an order of magnitude. Charge transfers in 2D heterobilayers, where electrons (holes) in one of the monolayers with a higher conduction band minima CBM (lower valence band maxima VBM) are transferred to the other monolayer, occur in the timescale of sub-picoseconds, regardless of the momentum mismatch between the layers.[9,10] Energy transfers in 2D heterobilayers, on the other hand, occur without net charge transfer between the layers, either by transfer of a set of electrons and holes (Dexter-type) or by a dipole coupling (Förster-type).[11,12] Various approaches have been taken to engineer the vdW interaction in the 2D materials system, such as thermal annealing,[13] surface termination,[14] intercalations,[15] interlayer spacing,[11,12] and electrical gating.[16,17] However, the most effective and chemically inert modulation of the interlayer interaction is through the application of mechanical strain.[18,19] A clear indication of the interaction modulation can be found in the reports for the band gap evolution of semiconducting TMDCs under hydrostatic pressure: Multi-layer semiconductor TMDCs turn metallic as applied pressure enhances the interlayer electronic interactions between sulfur and molybdenum atoms,[20] whereas the band gaps in monolayers open up with increasing pressure before it starts to close down at much higher pressures.[21,22] The enhanced vdW interaction can also lead to the modulation of the 2D materials' electronic,[23–25] thermal,[26,27] and topological[28,29] properties. Enhancement of the interlayer CTs has also been demonstrated in 2D heterostructure systems, clearly demonstrating the effect of the mechanical strain on the interlayer interaction.[30–32]

Here, we present an effective means of using compressive strain to tune the CT and excitonic ET in a $MoS_2/WS_2$ TMDC heterostructure under hydrostatic pressure. Vibrational properties, band gaps, and interlayer charge/energy transfers in the 1L-$MoS_2$/1L-$WS_2$ heterostructure were characterized by Raman and photoluminescence (PL) spectroscopy. At pressures below 4 GPa, the ET is dominant in the hetero-interface, where the PL intensity of $WS_2$ in the heterobilayer is strongly quenched compared to that of the stand-alone 1L-$WS_2$. Interestingly, CT starts to enhance greatly at pressures above 4 GPa, and the heterostructure-to-monolayer $WS_2$ PL intensity ratio increases more than an order of magnitude, whereas the ratio of $MoS_2$ continues to decrease. Density functional theory (DFT) calculations reveal that the drastic shift in dominant dynamics from the ET at lower pressure to CT at higher pressure is due to the enhanced vdW interaction and charge density evolution triggered by orbital switching in band extrema. The pressure effects are decomposed computationally to distinguish the effective strain component in the heterobilayer. The presented engineering of the vdW interaction and charge/energy transfers suggests a novel mechanism to potentially engineer mechanically-controlled optoelectronic devices.[33,34]

**Results**

*Sample Preparation and Initial Characterization*

High-quality single-crystal MoS$_2$ and WS$_2$ monolayers with lateral dimensions of approximately 100-400 μm were mechanically exfoliated using gold-assisted methods **(Figure 1a and 1b)**.[35] The MoS$_2$ and WS$_2$ monolayers were then stacked into a heterostructure (HS) on a 10 μm-thick single-crystal Si (100) wafer chip deposited with 25 nm Al$_2$O$_3$ using the dry-transfer method.[36] The HS on the substrate has long edges of the monolayers aligned in parallel with an overlapping area of ~50 μm × 70 μm **(Figure 1c)**. The large lateral dimensions of the heterostructure assured optical observations to be free of effects from possible edge states, grain boundaries, exciton diffusion, and flake-to-flake variations.[37,38] A ruby sphere was placed into the sample chamber, and high-purity Ne inert gas was used as the pressure transmitting medium (PTM) in order to exert quasi-hydrostatic pressure and to ensure chemical purity in the sample chamber. Note that Si and Al$_2$O$_3$ do not undergo first-order phase transitions in the pressure range of the present study, making a suitable substrate for the heterostructure.[39–41] Optical characterizations confirmed the uniform monolayer thickness and high quality of the exfoliated flakes in atmospheric pressure, as shown in the sharp Raman spectra **(Figure 1d)** and uniform Raman intensity maps of A′ peak in 1L-MoS$_2$ and 1L-WS$_2$ **(Figure S1)**. The Raman spectra of the heterostructure were a superimposition of individual MoS$_2$ and WS$_2$ spectra, with A′, E′, and 2LA(M) peaks present. Monolayer MoS$_2$ and WS$_2$ exhibit strong PL signals at the A exciton energies of 1.85 eV and 2.02 eV, respectively, in highly symmetric Gaussian lineshapes, confirming the high quality nature of the monolayers **(Figure 1e).** The PL spectrum of the heterostructure shows superimposed spectral signatures of both $MoS_2$ and $WS_2$. No interlayer excitons were observed in the present sample despite the parallel alignment of the monolayers.

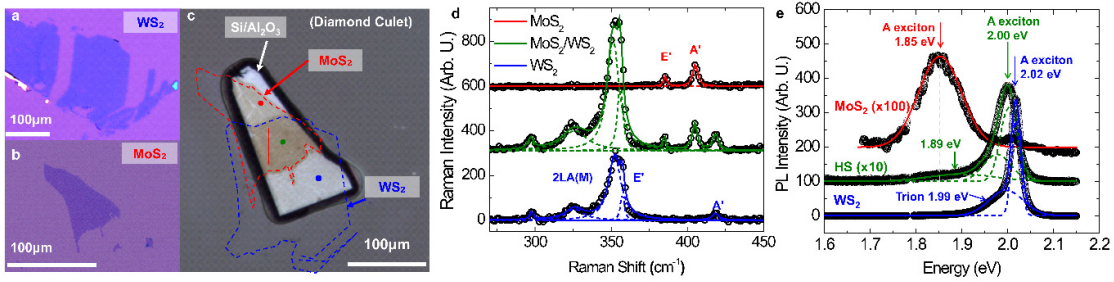

**Figure 1.** Preparation and Ambient-Pressure Characterization of the $MoS_2$/$WS_2$ Heterostructure Sample. Monolayer (a) $WS_2$, (b) $MoS_2$, and (c) their heterostructure (HS) transferred onto an alumina-coated silicon chip placed on a diamond culet. Red and blue dashed lines outline the $MoS_2$ and $WS_2$ monolayer, respectively, while their overlapping area represents the heterostructure. (d) Raman spectra of the monolayers and the HS, showing characteristic in-plane E′, out-of-plane A′, and second-order longitudinal acoustic LA(M) peaks. (e) photoluminescence (PL) spectra of the monolayers and HS. $MoS_2$, $WS_2$, and heterostructure data in (d) and (e) were measured from spots indicated as red, blue, and green dots marked in (c), respectively.

*Strain-Modulation of Phonon Dynamics and Band Structures*

With hydrostatic pressure applied to the sample using a diamond anvil cell (DAC) apparatus **(Figure 2a)**, the in-plane E′ and out-of-plane A′ Raman mode **(Figure 2b-c)** frequencies of the 1L-$MoS_2$, 1L-$WS_2$, and their HS blueshift, an indication of the compression-induced bond length shortening (symbols in **Figure 2d-e** and **Figure S2**). The A′ peaks display significantly higher pressure-dependence (≈3.3 cm$^{−1}$ GPa$^{−1}$) than that of the E′ peaks (≈2.3 cm$^{−1}$ GPa$^{−1}$), showing that the hydrostatic pressure is effective in shortening the interlayer distance and modulating the out-of-plane modes. The E′ peaks in the HS show insignificant deviation from those in the monolayers, confirming minimal slippage at the hetero-interface. On the other hand, the A′ peaks in the heterostructures are consistently lower than those in the monolayers **(inset of Figure 2e)**, which is an indication of the enhanced interlayer interaction.[20,30] The weaker blueshifts observed here contradict a previous report on $MoSe_2$/$WSe_2$ heterostructure[32], where the discrepancy may arise from differences in the material combination or experimental setups. To computationally model the effect of hydrostatic pressure on the vibrational mode evolution, Raman modes were calculated for the monolayers and the HS, under in-plane strain, out-of-plane strain (lines in **Figure 2d-e**), and hydrostatic pressure conditions (**Figure S5-6** and **Table S1**). All Raman active modes of monolayers show significant blueshifts with increasing pressure, in good agreement with the experimental observation. For decomposed strain components, it is found that in-plane (dashed lines) and out-of-plane (solid lines) strain components have a similar contribution to the blueshift of the E′ modes, whereas the out-of-plane strain component contributes the most in the A′ mode blueshift. These experimental and computational findings highlight the role of hydrostatic pressure in enhancing the out-of-plane interlayer interactions. Details on the computation of active Raman modes (**Figure S3**) and strain components (**Note S1, Figure S4**) could be found in the Supporting Information.

The optical band gaps of the monolayers and HS are extracted from the Gaussian fitting of the PL spectra at high pressures (**Figure S7**). The band gaps of the $MoS_2$ in both monolayer and heterostructure increase with increasing pressure with a nearly linear pressure dependence of ~25.3 meV $GPa^{-1}$, in good agreement with previous reports (**Figure 3a**).[21,22] On the other hand, the band gaps of the $WS_2$ increase until reaching an extremum at ~4 GPa and then start to moderately decrease with further increase in pressure. The non-monotonic pressure response of the $WS_2$ band gap can be attributed to the transition from the K-K direct bandgap to the Λ-K indirect bandgap.[42] Our theoretical calculations confirm the decomposition of the strain-induced band gap modulation as well. The evolution of the excitonic energy levels with different strain components shows that the in-plane strain component is most effective in increasing the band gaps (dashed lines in **Figure 3c-d** and **Figure S8, 9**), whereas the out-of-plane strain component (dotted lines) suppresses the band gaps. As a result of these competing effects, the hydrostatic pressure increases the band gap, which is in good agreement with our experimental results and previous reports.[21] The band gap decrease from the out-of-plane strain component is more prominent in the HS than in the monolayers, further supporting the notion that the interlayer interaction plays an important role in the band gap closing **(Table S2)**. The strong band gap opening effect implies that in-plane strain dominates the band structure modulation and is responsible for the indirect band gap transition at higher pressures. Based on our electronic band gap calculations **(Figure S10)**, out-of-plane strain increases the valence and conduction band offsets and therefore decreases the indirect band gap.

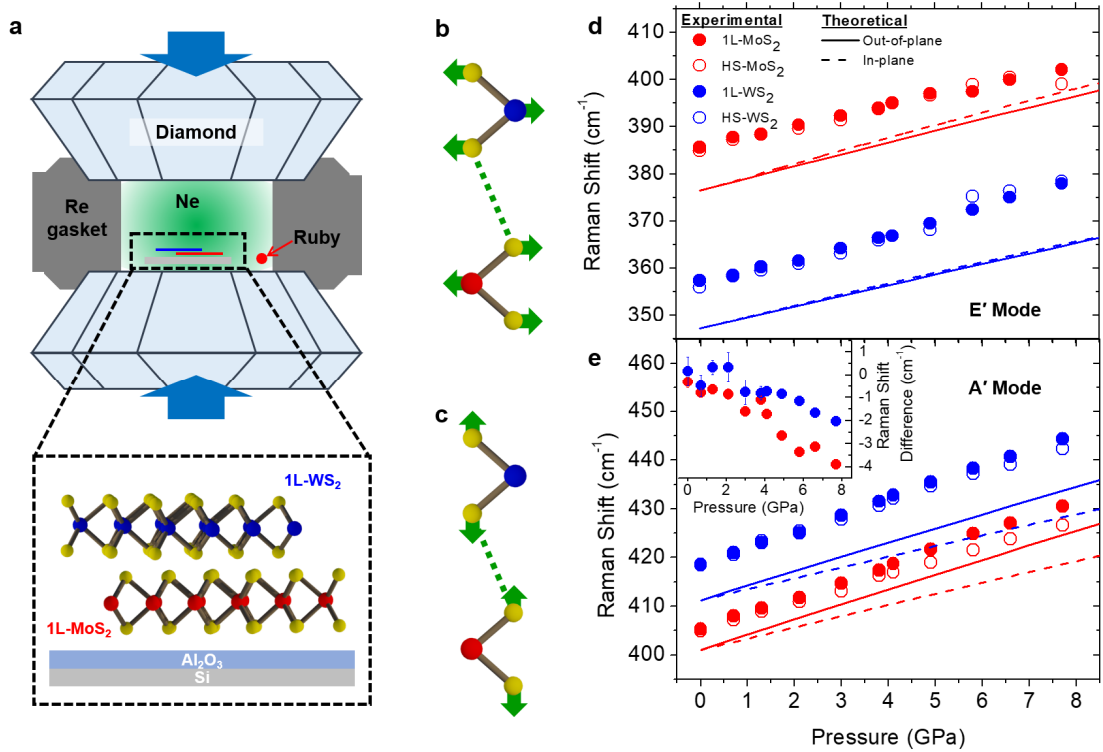

**Figure 2.** Raman Measurements of the 1L-MoS$_2$/1L-WS$_2$ Heterostructure under Hydrostatic Pressure. (a) Schematic side view of the sample chamber in a diamond anvil cell (top) and the HS sample on Al$_2$O$_3$/Si substrate (bottom). (b) In-plane E′ and (c) out-of-plane A′ Raman modes with dashed lines illustrating the interlayer interaction. (d) In-plane E′ and (e) out-of-plane A′ Raman peak positions from MoS$_2$ and WS$_2$, experimentally measured (symbols) and theoretically calculated (lines). Inset of (e) shows the peak position difference of the A′ Raman mode ($\omega_{HS} - \omega_{1L}$).

*Strain-Modulation of Charge and Energy Transfers*

In order to decipher the pressure-evolution of the interlayer interaction in the heterostructure, the enhancement factor η, defined as the experimentally measured PL intensity ratio between the HS and the monolayers ($I_{HS}/I_{1L}$), was extracted as a function of pressure **(Figure 3b)**. At pressures below ~4 GPa, the enhancement factor of the MoS$_2$ remains close to unity with a marginal decrease. On the other hand, $\eta_{WS2}$ is significantly lower than unity at ~0.7 GPa and drastically decreases with pressure. The asymmetric PL quenching in the heterostructure indicates that the ET is the dominating charge dynamics across the vdW interface.[43] While the $\eta_{MoS2}$ ratio continues to decrease, the PL emission landscape of WS$_2$ changes above 4 GPa: the $\eta_{WS2}$ begins to increase, exceeds that of lower-pressure values, and eventually reaches unity at 7.7 GPa. The increased $\eta_{WS2}$ suggests that the interlayer ET has subsided, and CT has become dominant. The anomalous evolution of the enhancement factors implies strong interlayer interactions under pressure, which lead to the modulation of the charge and energy transfers occurring at the heterostructure interface.

DFT band structure calculations were carried out to analyze the orbital contributions in the CBM and VBM **(Figure S11)**. The intralayer A excitons ($A_{MoS2}$ and $A_{WS2}$) of the monolayers are localized at the K-point in the momentum space of the heterostructure, as CBM (CB1) and second valence band (VB2) come from MoS$_2$, and VBM (VB1) and second conduction band (CB2) come from WS$_2$. The evolution of charge densities of the conduction and valence band valleys with the in-plane strain component is shown in **Figure 4a** and **4b**, respectively. From ambient pressure to ~13 GPa (P1), the contribution of WS$_2$ electronic states in CB2 and VB1 valley is less than that of the MoS$_2$ electronic states in CB1 and VB2 valley, suggesting that both electron and hole densities are lower in WS$_2$. The asymmetric charge density shows an energy transfer from WS$_2$ to MoS$_2$, supporting our experimental findings (schematic view shown in **Figure 4c**). On further increasing strain up to ~22 GPa (P2), the electron density of the CB2 valley (WS$_2$) decreases rapidly, implying an even larger energy transfer from WS$_2$ to MoS$_2$, which further explains the dramatic decrease of the WS$_2$ enhancement factor in experimental results.

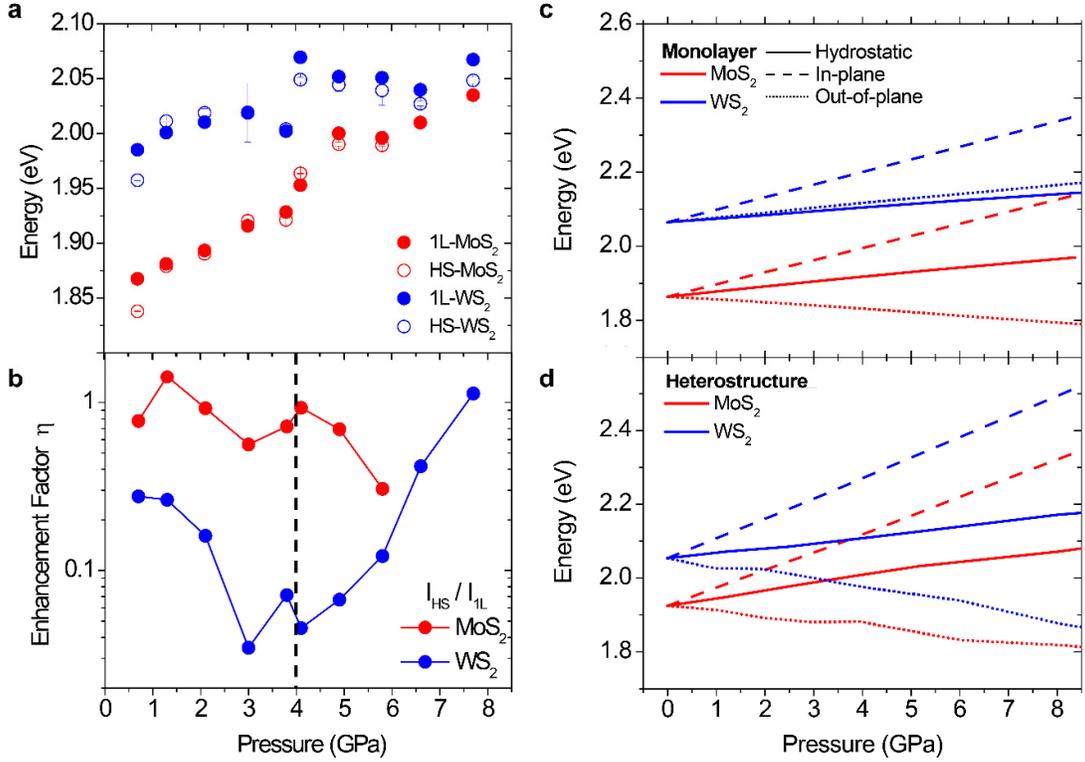

**Figure 3.** Pressure-Dependent Evolution of Optical Band Gaps and PL Intensity Ratio. (a) Optical band gaps of the monolayers and heterostructure measured from the PL spectral peaks as a function of hydrostatic pressure. (b) PL enhancement factor η ($I_{HS}/I_{1L}$) as a function of pressure. (c-d) Calculated evolution of the A exciton states of (c) individual monolayers and (d) as the heterostructure as a function of pressure (solid lines), and equivalent in-plane (dashed lines) and out-of-plane strains (dotted lines).

The charge contributions in CB1 and CB2 valleys of the heterostructure band are further analyzed in terms of the charge redistribution, as shown in **Figure 4d**. In the unstrained case, the CB1 (lower panel) and CB2 (upper panel) originate from the major contribution of $d_{z^2}$ orbitals of Mo (blue) and W (yellow) atoms, respectively. With increasing pressure along the x-y plane, the W-W and Mo-Mo atoms and their $d_{z^2}$ orbitals come closer to each other **(Figure S12)** until the ring of each $d_{z^2}$ orbitals overlaps near the critical transition points of the charge transfer, e.g., at P1 point for CB2 and P2 point for CB1, respectively. The higher onset pressure for the charge transfer in the CB1 ($MoS_2$) valley is attributed to the less diffusive nature of Mo $d_{z^2}$ orbitals. The charge transfer rate is also lower for CB1 compared to CB2 valley, which is also experimentally reflected in the smaller change in the PL enhancement factor. After increasing pressure from P2 up to ~30 GPa (P3 point), the rest of the charges in the CB2 valley transfer from $WS_2$ to $MoS_2$ monolayer, which is responsible for the further decrease in the $A_{WS2}$ and $A_{MoS2}$ peak intensity in the heterostructure. However, above the P3 pressure, the $\eta_{WS2}$ starts increasing again due to the increase in the population of $WS_2$ states at CB2 valley. On the other hand, the $\eta_{MoS2}$ continues to decrease due to decreased $MoS_2$ states in CB1 valley.

The abrupt changes in the charge densities are investigated from the molecular orbital (MO) diagram at the K point for different strains, as shown in **Note S2** and **Figure S13**.[44] At ~22 GPa pressure (above P1), the orbital crossing happens in the vdW heterostructure between the CB2 and CB3 valleys, as shown in **Figure 4d**. Similarly, by applying ~30 GPa pressure (above P2), the orbital switching occurs between CB1/CB2, CB3/CB4, and VB3/VB4 valleys. At ~39 GPa pressure (above P3), CB2/CB3 orbital crossing takes place. These orbital switchings in individual layers are responsible for the drastic turnovers in the energy transfer and charge transfer in the $MoS_2/WS_2$ heterostructure. Most importantly, the out-of-plane strain is responsible for enhancing the interlayer orbital coupling, thereby greatly accelerating the strain-dependent changes in the orbital contributions and charge density.[20]

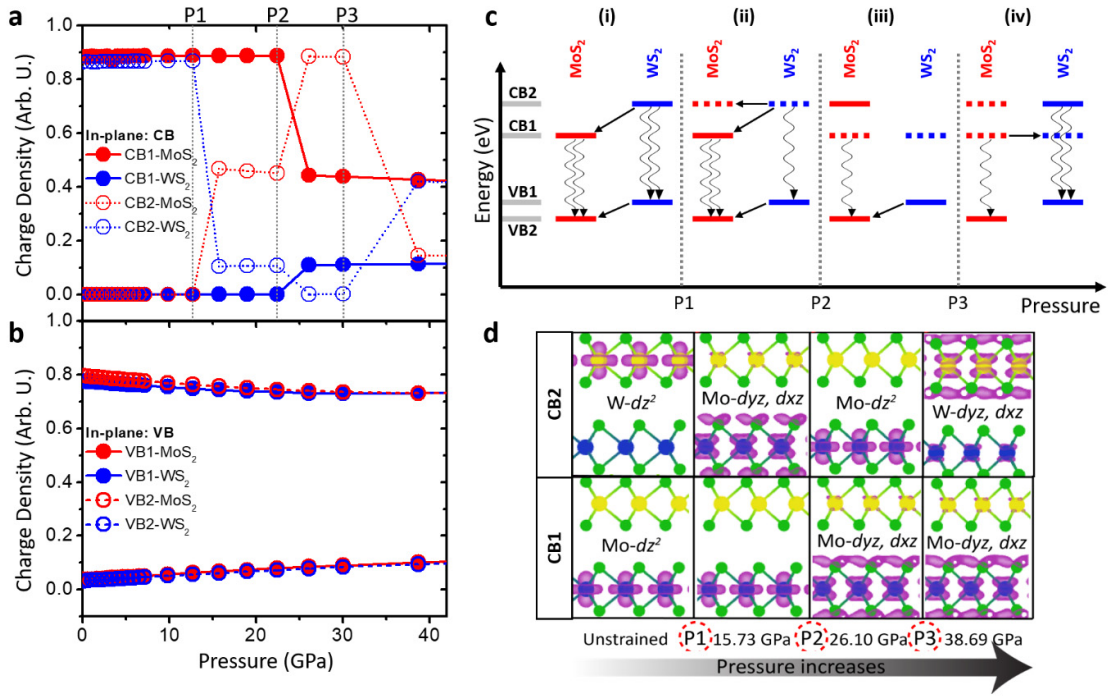

**Figure 4.** Pressure-Dependent Charge and Energy Transfers of the $MoS_2/WS_2$ Heterobilayer. (a-b) Charge density at (a) conduction band and (b) valence band as a function of pressure along the in-plane direction. (c) Schematics demonstrating the charge density and charge transfers with increasing pressure. Straight and wavy arrows indicate the charge transfer and radiative decay, respectively. (d) Band-decomposed charge densities of the $MoS_2/WS_2$ heterostructure, superimposed with the side view of $MoS_2$ (bottom) / $WS_2$ (top) heterostructure. P1, P2, and P3 are representative pressure points where significant changes occur in the charge or orbital contribution to the conduction band, which correspond to ~13 GPa, ~22 GPa, and ~30 GPa, respectively.

In summary, compressive strain under hydrostatic pressure has been applied to $MoS_2/WS_2$ heterobilayer to modulate their interlayer interactions and to control the interlayer charge and energy transfers. Enhanced interlayer interaction due to out-of-plane compressive strain was observed in blueshifts in Raman spectroscopy and photoluminescence spectroscopy, and Density Functional Theory was carried out to investigate the strain components' contributions. The out-of-plane strain component effectively blueshifts A′ Raman modes and decreases the optical band gap, indicative of the strong interlayer interaction. PL enhancement factor for $WS_2$ decreases by the factor of five up to 4 GPa, but substantially increases by an order of magnitude up to 7.7 GPa, which is attributed to the enhanced energy transfer and charge transfer, respectively. DFT calculations reveal that the strong enhancement and modulation of the energy and charge transfers are a result of the orbital switching and charge density change in the heterostructure's conduction and valence band extrema. Our findings highlight an effective way of controlling the interlayer energy and charge transfer in the 2D heterobilayer by compressive strain, thereby opening opportunities to mechanically modulate excitonic device functionalities.

**Methods**

*Preparation of the Heterobilayer Sample*

Bulk $MoS_2$ and $WS_2$ crystals (2D Semiconductors) were mechanically exfoliated onto $Si/SiO_2$ substrates, and then evaporated with 50 nm-thick gold at a rate of 1 Å s$^{-1}$ using an e-beam evaporator. A thermal release tape was then attached and peeled to isolate the topmost monolayers and the Au film.[35] The monolayer-Au films were then thermally released on desired target substrates ($SiO_2/Si$), where chemically etching the gold film using potassium iodide/iodine ($KI/I_2$) solution resulted in single-crystal monolayers with 100-400 μm lateral size. Finally, the monolayers were rinsed with acetone and isopropyl alcohol (IPA) to remove any remaining residues. For stacking and transferring the heterobilayer, a PDMS stamp with a PPC adhesion layer was used to first pick up the $WS_2$ monolayer from the silicon substrate. The $WS_2$ monolayer was aligned and contacted with the $MoS_2$ monolayer, and subsequently used to pick up the $MoS_2$ monolayer with a precision alignment. The picked-up heterobilayer was dropped onto the target substrate ($Al_2O_3/Si$ placed on a diamond culet) by thermally melting the PPC. The PPC was eventually dissolved using acetone/isopropanol mixture.

*Sample preparation for diamond anvil cell experiments*

A pair of 400 μm culet diamond anvils with ultralow fluorescence backgrounds were used in a symmetric DAC for the experiments. A Re gasket was pre-indented to ~45 μm thick, and subsequently a ~230 μm diameter hole was drilled at the very center of the indentation to form a sample chamber. A 10 μm-thick polished single-crystal Si chip (University Wafers) was deposited with 25 nm-thick $Al_2O_3$ using atomic layer deposition (ALD), cleaved into a desirable size, and placed onto one of the culets. After transferring the heterobilayer sample on the $Si/Al_2O_3$ substrate, ultrahigh purity Ne pressure medium (99.999%, Airgas) was loaded into the sample chamber using a home-built gas loading system in the Mineral Physics Laboratory at The University of Texas at Austin. The pressure before and after each spectroscopic measurement was monitored by measuring the $R_1$ fluorescence line of the ruby sphere placed near the sample.[45]

*Optical characterization*

Raman and PL spectroscopic measurements were carried out using the Renishaw inVia micro-Raman system at the Mineral Physical Laboratory at the University of Texas at Austin. A focused laser beam with a wavelength of 532 nm, spot size of ≈1 μm and incident power <5 mW at the sample position was used for the measurements. The spectral resolution in Raman measurements is ≈1.3 $cm^{-1}$. Lorentzian and Gaussian fittings were carried out to extract peak parameters from Raman and PL spectra, respectively.

*Theoretical Calculations*

Theoretical calculations of the monolayers and heterostructure of TMDCs were performed using first-principle density functional theory (DFT) as implemented in Vienna Ab-initio Simulation Package (VASP) [46,47]. All-electron projector augmented wave (PAW) potentials[48,49] were used to describe electron-ion interactions. The electronic exchange and correlation potential terms were represented by the Perdew-Burke-Ernzerhof (PBE)[50] generalized gradient approximation (GGA). The Kohn-Sham orbitals were expanded in the plane wave basis sets with an energy cutoff of 500 eV. All monolayers and heterostructure were relaxed using the conjugate-gradient method until the Hellmann-Feynman forces on every atom were less than 0.005 eV Å$^{-1}$. A well converged Monkhorst-Pack (MP)[51] k-grid of 12×12×1 was used to sample the Brillouin zone (BZ). To avoid spurious interactions between the periodically repeated images, a vacuum of 20 Å was used along the z-axis. The weak van der Waal (vdW) interaction between the layers was incorporated by the Grimme's PBE-D2 exchange functional.[52] The Raman modes of monolayers and heterostructure of TMDCs were calculated by using the density functional perturbation theory (DFPT).[53] To obtain accurate quasiparticle band structure and the correct description of the optical properties of the monolayers and heterostructure of TMDCs, many-body interactions should be taken into account in the calculations. The single shot GW (i.e., G0W0) calculation,[54] based on the many-body perturbation theory, was performed to get the quasiparticle information. Finally, we carried out the Bethe-Salpeter equation (BSE) calculations on top of G0W0 in order to get the information of excitonic effects in the optical absorption using the Tamm-Dancoff approximation.[55] For GW calculations, a k-mesh of 12×12×1 within the MP scheme was used for monolayers and heterostructure system. The energy cutoff for the wave functions and response functions were set to be 500 eV and 200 eV, respectively, in BSE calculations. A well converged 432 empty bands and 50 frequency grid points were taken in the calculations. The 9 highest occupied valence bands and 18 lowest unoccupied conduction bands were included for excitonic states.

**Data Availability**

The data supporting the findings of this work are available within the paper and supplementary information, as well as from the corresponding authors upon reasonable request.

**Acknowledgements**

J.-S.K. and D.A. acknowledge support from the Defense Threat Reduction Agency (DTRA). J.-S.K. acknowledge support from NSF MRSEC program (DMR-1720139), and M.K. from NSF NASCENT ERC Center, respectively. N.M., R.J., and A.K.S. thank Materials Research Centre and Thematic Unit of Excellence, Indian Institute of Science, for providing the computational facilities. N.M., R.J., and A.K.S. acknowledge the support from Institute of Eminence (IoE) MHRD grant of Indian Institute of Science, and the grant from DST Korea. J.F.L. acknowledges support for the Renishaw inVia Raman system from Department of Geological Sciences and Jackson School of Geosciences at the University of Texas at Austin.

**Author Contributions**

J.-S.K, D.A, and J.-F.L. conceived the original idea for the study. J.-S.K. designed the experiments, performed pressure-dependent Raman and PL measurements, and analyzed the data. M.K. provided large-scale exfoliated 2D materials, and S.F. helped to load neon pressure medium to the diamond anvil cell. All experimental work was conducted at the University of Texas at Austin. N.M. and A.K.S. designed the theoretical calculations and corresponding analysis. N.M. and R.J. conducted theoretical calculations. A.K.S, D.A, and J.-F.L. coordinated and supervised the research. All authors contributed to the article based on the draft written by J.-S.K.

**Corresponding Authors**

Correspondence to Joon-Seok Kim, Deji Akinwande, and Jung-Fu Lin